# Twisted toroidal vortex-solitons in inhomogeneous media with repulsive nonlinearity

Y. V. Kartashov,[1,2] B. A. Malomed,[3] Y. Shnir,[4,5] and L. Torner[1]

[1]ICFO-Institut de Ciencies Fotoniques, and Universitat Politecnica de Catalunya, Mediterranean Technology Park, E-08860 Castelldefels (Barcelona), Spain
[2]Institute of Spectroscopy, Russian Academy of Sciences, Troitsk, 142190 Moscow, Russia
[3]Department of Physical Electronics, School of Electrical Engineering, Faculty of Engineering, Tel Aviv University, 69978 Tel Aviv, Israel
[4]Bogoliubov Laboratory for Theoretical Physics, Joint Institute for Nuclear Research, Dubna 141980, Moscow Region, Russia
[5]Institute of Physics, Carl von Ossietzky University Oldenburg, Oldenburg, D-26111, Germany

Toroidal modes in the form of so-called Hopfions, with two independent winding numbers, a hidden one (twist, $s$), which characterizes a circular vortex thread embedded into a three-dimensional soliton, and the vorticity around the vertical axis $(m)$, appear in many fields, including the field theory, ferromagnetics, and semi- and superconductors. Such topological states are normally generated in multi-component systems, or as trapped quasi-linear modes in toroidal potentials. We uncover that stable solitons with this structure can be created, without any linear potential, in the *single-component* setting with the strength of repulsive nonlinearity growing fast enough from the center to the periphery, for both steep and smooth modulation profiles. Toroidal modes with $s=1$ and vorticity $m=0,1,2$ are produced. They are stable for $m \leq 1$, and do not exist for $s>1$ An approximate analytical solution is obtained for the twisted ring with $s=1, m=0$. Under the application of an external torque, it rotates like a solid ring. The setting can be implemented in BEC, by means of the Feshbach resonance controlled by inhomogeneous magnetic fields.

A fundamental topic in many fields of physics is the creation of three-dimensional (3D) self-trapped nonlinear modes (solitons). Ramifications of this topic are well known in optics [1], Bose-Einstein condensates (BECs) [2,3], ferromagnetics [4], superconductors [5], semiconductors [6], nuclear matter [7], and the field theory [8-10]. Self-attractive nonlinearity is usually needed for the formation of localized states. This causes a major problem, as attractive cubic nonlinearities cause collapse of multi-dimensional states [11] and azimuthal instabilities of ring-shaped vortices [12].

Fundamental and vortical 3D solitons can be stabilized by lattice potentials [1,13]. 3D objects may also be stable in nonlocal nonlinear media [14]. Spin-orbit interactions in BECs may stabilize 2D solitons in free space [15]. On the other hand, nonlinear *pseudo-potentials*, induced by periodic modulation of the local strength of the nonlinearity, do not stabilize 3D solitons. Stabilization of 2D states has been shown in pseudo-potentials whose shapes feature sharp edges [16].

A completely different approach to the problem was proposed in Refs. [17-19], where it was shown that *repulsive* spatially inhomogeneous nonlinearity, with the local strength, $\sigma(\rho)$, growing as a function of radial variable $\rho$ faster than $\rho^3$, creates stable fundamental- and vortex-soliton states. In BEC, the required spatial modulation of the nonlinearity strength may be induced by means of suitable Feshbach resonances (FRs) [20-23] controlled by inhomogeneous magnetic [24-26] or laser [27] fields (necessary physical conditions for that are considered below).

In 3D geometry, a challenge is to construct stable vortex-soliton states with complex structures, such as Skyrmions and Hopfions, which carry *two independent winding numbers*. The aim of this Letter is to show that an apparently simple isotropic model with a single wavefunction generates 3D solitons in the form of stable vortex rings with *internal twist*. For these solitons, the phase of the wavefunction changes both along and around the vortex ring, with the corresponding topological invariant ( linking number) is the product of the number of twists ($s$) and overall vorticity ($m$). Since this arrangement is typical for solitons of the Faddeev-Skyrme model (FSM) [10,28-30], with the triplet of real scalar fields realizing the Hopf map, $\varphi: R^3 \to S^2$, such states are called *Hopfions*. So far, Hopfions were found only in systems with multicomponent wavefunctions [8,31]. Here we show, for the first time to our knowledge, that the usual matter-wave model, based on the *single* Gross-Pitaevskii equation (GPE), with $\sigma(\rho)$ growing as said above, gives rise to several species (including stable ones) of twisted vortex rings with two independent winding numbers.

The GPE for the scaled wave function, $q(x,y,z,t)$, is written in the Cartesian coordinates, $x,y,z$, and normalized time, $t$:

$$iq_t = -(1/2))(q_{xx} + q_{yy} + q_{zz}) + \sigma(\rho)|q|^2 q, \qquad (1)$$

where $\rho^2 \equiv x^2 + y^2 + z^2$, and the local strength of the repulsive nonlinearity, $\sigma(\rho) > 0$, must grow faster than $\rho^3$, to secure the convergence of the norm, $N = \iiint |q|^2 dxdydz$.

The required FR-modulation profile may be imposed by a magnetic pattern which provides for the exact FR at $\rho \to \infty$ (in fact, at a relatively large finite value of $\rho$, which is sufficient as the respective solitons are strongly localized [17-19]), e.g., $H_{\text{FR}} = 402$ G and $735$ G for $^{39}$K [20] and $^7$Li [21], respectively. Using crossed fields induced by barrel-shaped solenoids with trapped magnetic flux $\Phi$, whose radii vary along $\rho$ (realized here as the coordinate running along the solenoid's axis) as $R^2 = 3^{1/2}\Phi/(\pi H_{\text{FR}}) + \text{const}/\sigma(\rho)$ [32], one may construct the required approximately isotropic pattern. It can be created in a more accurate form by means of recently developed techniques, *viz.*, magnetic lattices [24-26], *field concentrators* [33], and current circuitry integrated with the trap [34]. An estimate for the $^7$Li BEC, where the FR was studied in detail [21], suggests that the present system may be realized with magnetic-field gradients $\sim 10$ T/m. Very recent work demonstrates that so high gradients become available [34].

Here we present numerical and analytical results for the model with a steep anti-Gaussian modulation, $\sigma = \exp(\rho^2/2)$, and for a smooth power-law profile, $\sigma = 1 + \rho^6$. The former choice makes it possible to display the results in the most compact form, while the latter one demonstrates that the findings are generic, being valid for all modulation formats subject to the above-mentioned condition, $\rho^3/\sigma(\rho) \to 0$ at $\rho \to \infty$. Note that the vanishing of the wave function at $\rho \to \infty$ provides for a possibility of the one-point compactification of the boundary, $R^3 \to S^3$. Then, the subsequent map onto the target space represents the *Hopf fibration* [35] in the present setting.

Because Eq. (1) does not include any trapping potential, all the self-trapped modes presented below are nonlinear states, which do not bifurcate from eigenmodes of any linear model. In this connection, it is relevant to mention that twisted toroidal vortices may exist as linear modes in torus-shaped trapping potentials. In the latter case, the addition of repulsive nonlinearity leads to deformation of the trapped vortical rings, without a qualitative change in their shape [36].

Using cylindrical coordinates $(r,\phi,z)$, we seek for stationary solutions of Eq. (1) with integer vorticity $m$ and chemical potential $\mu$ as $q = w(r,z)\exp(im\phi)\exp(-i\mu t)$, where the amplitude function obeys the stationary GPE:

$$\mu w = -(1/2)(w_{rr} + r^{-1}w_r - m^2 r^{-2}w + w_{zz})w + \sigma(r,z)|w|^2 w. \qquad (2)$$

In contrast with the usual vortex rings featuring the single winding number, $m$, in the $(x,y)$ plane, here the *complex* amplitude function, $w(r,z) = w_r(r,z) + iw_i(r,z)$, accounts for the *second winding number* (twist), $s$, in the $(r,z)$ plane. This implies a structure in the form of a toroidal *twisted vortex tube*, nested in the 3D soliton and coiling up around the $z$ axis, see examples in Figs. 1 and 2. Angular momentum $M$ of such states is linked to the norm as for usual vortices with $s = 0$, i.e., $M = mN$.

Counterparts of $w_r$ and $w_i$ in the FSM are two components of the triplet of real fields which is restricted to the surface of the unit sphere [10,28,29], and the twisted rings correspond to the fundamental knots (the so-called *unknots*) [37]. The present model also bears certain similarity to the *twisted Q-balls*, i.e., stationary rotating non-topological solitons in field models with polynomial nonlinearity [10,29].

Numerical solutions of Eq. (2) were obtained by the relaxation method. Examples of stationary twisted toroidal states with winding number $s=1$, $m=0,1,2$ are displayed in Fig. 1 for the model with the steep nonlinearity modulation, while solitons with $s=1$, $m=0$ in the model with the smooth modulation are shown in Fig. 2. To the best of our knowledge, Hopfions shown in Figs. 1 and 2 represent the first examples of the toroidal twisted vortex tube, nested in localized modes, which self-trap in a scalar model. The coiled vortex thread and the usual straight vortex axis (for $m=1,2$) are clearly visible in Figs. 1 and 2. No solutions have been found for $s>1$, suggesting that the single nonzero value of the intrinsic winding number may be $s=1$.

The main properties of the toroidal vortex-soliton families are summarized in Fig. 3. They satisfy the *anti-Vakhitov-Kolokolov* criterion, $dN/d\mu > 0$, which is a necessary stability condition for localized modes supported by repulsive nonlinearities [38]. The state with smaller $m$ always possesses a larger norm. Figure 3(c) shows that the radius of the coiled vortex thread decreases with the increase of norm $N$, remaining *finite* (not collapsing to zero) at $N \to \infty$.

Physical parameters of these modes can be estimated using the above-mentioned data for $^7$Li [21]. The result is that the twisted rings of radius $\sim 10~\mu m$, with the density peaked in the region with the local value of the scattering length $\sim 5$ nm, can be built of $\sim 10^3$ atoms, the respective time scale being $\sim 10$ ms.

Analytical results can be obtained by means of the variational approximation (VA), based on the Lagrangian of Eq. (2), $L = \pi \int_0^\infty r dr \int_{-\infty}^\infty dz \{ 2[(m/r)^2 - \mu]|w|^2 + |\nabla w|^2 + \sigma |w|^2 \}$. In the model with the steep modulation, we use a natural ansatz, $w(r,z) = A[(r-R)+iz]^s r^m e^{-(r^2+z^2)/2}$, for the wave function with two winding numbers, where $A$ and $R$ are the soliton's amplitude and radius of the coiled vortex axis. Results are presented here for the basic twisted toroidal state with $m=0$.

The norm of the ansatz is $N = (2\pi)^{3/2} dA^2$, with $d \equiv R^2 - (2\pi)^{1/2}R + 3$. Substitution of the ansatz into the Lagrangian yields $L_{eff} = -\mu N + (N/8d)[3R^2 - (8\pi)^{1/2}R + 11] + (N^2/2d^2(2\pi)^{3/2})[R^4 - (8\pi)^{1/2}R^3 + 14R^2 - 8(2\pi)^{1/2}R + 15]$. The variational equation following from here, $\partial L_{eff}/\partial R = 0$, yields the radius of the coiled vortical axis as a function of the norm. In the Thomas-Fermi (TF) limit, $N \to \infty$, this equation takes the form of $2(4-\pi)R_\infty^3 - 3(2\pi)^{1/2}R_\infty^2 + 4(2\pi-3)R_\infty = 3(2\pi)^{1/2}$, yielding a finite radius, $R_\infty \approx 1.07$. In fact, $R_\infty$ is an *asymptotically exact* value of the radius in the limit of $N \to \infty$. The modal width, defined by $W^2 = 4N^{-1} \iiint (x^2+y^2+z^2)|q(x,y,z)|^2 dxdydz$, also attains an asymptotically constant value at $N \to \infty$, which is predicted by the TF approximation (cf. Ref. [18]): $W_{TF}^2 = 4 \iint \rho^2 d\rho/\sigma(\rho) / \int \rho^4 d\rho/\sigma(\rho)$. As seen in Fig. 3(b), these values for the steep and smooth modulations, $W_{TF}^2 = 12$ and $8$, respectively, are close to their numerical counterparts. On the other hand, at $N \to 0$ the toroidal modes expand indefinitely, corroborating the conclusion that they do not bifurcate from linear eigenmodes.

In Fig. 3(c), the VA shows close agreement with numerical findings, starting from moderate norms, $N \sim 100$, and confirm that the above value $R_\infty$ is asymptotically exact for $N \to \infty$ and $m=0$. The TF approximation can be used to estimate the radius of the vortex core surrounding the coiled vortical axis in the $(r,z)$ plane, $\rho_0 \approx (2\pi^3)^{1/4} N^{-1/2}$. It shrinks at $N \to \infty$ under the action of the nonlinear pressure.

The stability of the twisted toroidal modes was tested by direct simulations, and also by means of the linear-stability analysis. For the latter purpose, a perturbed solution is substituted into Eq. (1) as $q = (w + u e^{ik\phi + \delta t} + v^* e^{-ik\phi + \delta^* t}) e^{im\phi - i\mu t}$, where $u, v$ are small perturbations with integer azimuthal index $k$ and growth rate $\delta$. The resulting linearized eigenvalue problem was solved numerically. We have found that the basic twisted toroidal state, with $s=1, m=0$, is *completely stable*, for the steep and smooth modulation formats alike. An example of the corresponding stable evolution in the former case is displayed in Fig. 4. The vortices with $s=1, m=1$ are unstable at $\mu < \mu_{min} = 11.2$, and stable at $\mu > \mu_{min}$. The most destructive perturbations chiefly correspond to azimuthal in-dex $k=1$, as shown in Fig. 3(d), while $k=2$ is dominant at $\mu \to 0$. The development of the instability causes bending of the straight vortex line, which eventually touches the circular vortex thread, destroying the entire structure, as seen in Fig. 4. Higher-order vortices with $s=1, m \geq 2$ are completely unstable. The instability of the $m=2$ mode splits the straight vortex line into two, which then collide with the circular vortex thread, see Fig. 4.

The shape of the twisted toroidal state with $s=1, m=0$ suggests a possibility to set it in rotation around its diameter, like a solid ring, by applying an angular momentum in the plane of the ring. To this end, the wave function of the toroidal state was multiplied by $\exp[i\alpha z \tanh(x/x_0)]$, with constants $\alpha$ and $x_0$, which imparts angular momentum directed along $y$. As shown in Fig. 5, the ring-shaped state indeed responds by persistent rotation around the $y$ axis.

In summary, we have shown that 3D physical settings described by the GPE with the strength of the repulsive nonlinearity growing fast enough towards the periphery gives rise to robust states of the Hopfion type, i.e., twisted toroidal tubes with two independent winding numbers, the intrinsic one, $s$, and the conventional vorticity, $m$. This is the first example of a *single-component* model which produces such complex states in the absence of trapping potentials. The modes exist for $s=1$, being stable at $m \leq 1$. Steep and smooth nonlinearity-modulation profiles produce similar results. A challenging question, suggested by the analogy with the FSM [28], is whether the same model admits still more complex states, such as linked vortices and toroidal knots.

# Figure captions

Figure 1. (Color online) Left, central, and right columns: "Hopfions", with chemical potential $\mu=15$, intrinsic winding number $s=1$, and vorticity $m=0,1,2$, in the model with the steep modulation format. The first row shows the density surface at $|q(x,y,z)|^2=1$, the second and third rows display the absolute value and phase of the wave function in cross section $y=0$, and the fourth row shows the distribution of $|q(x,y,z)|$ at $z=0$. Here and below, blue and red regions in density plots correspond to lower and higher values, respectively.

Figure 2. (Color online) The same as in the three top rows in Fig. 1 for a stable "Hopfion" with $s=1$, $m=0$, and $\mu=10$, supported by the smooth modulation format. The density surface corresponds to $|q(x,y,z)|^2=0.4$.

Figure 3. (Color online) Families of twisted toroidal states with intrinsic winding number $s=1$ and different values of vorticity $m$: (a) the norm versus the chemical potential; (b) the width versus the norm; (c) the radius of the coiled vortex line, $R$, versus the norm [the blue curve in (c) shows $3N$]. Stable and unstable branches are depicted by black and red segments, respectively. Circles in (a) correspond to the modes displayed in Fig. 1. The chain of green dots in (c) depicts the VA prediction for $R$. Lower blue curves in (b,c) correspond to the states with $s=1, m=0$ in the model with the smooth modulation format, which are completely stable, while all other results pertain to the steep format. (d) The real part of the instability growth rate, for different values of azimuthal perturbation index $k$, versus the chemical potential, for $m=1$.

Figure 4. (Color online) Isosurface plots showing the perturbed evolution of the states with $\mu=15$, $s=1, m=0$ (the first row), $\mu=9.5$, $s=m=1$ (the second row), $\mu=16$, $s=m=1$ (the third row), and $\mu=22$, $s=1, m=2$ (the fourth row) in the model with the steep modulation. Surfaces are drawn at $|q(x,y,z)|^2=0.2$ in the second row, and at $|q(x,y,z)|^2=1$ in all others.

Figure 5. (Color online) Isosurface plots illustrating the solid-ring-like rotation of the twisted toroidal state with $s=1, m=0$, $\mu=10$, in the steep-modulation model, around its diameter (the $y$ axis), initiated by the torque with $\alpha=5$ and $x_0=5$ (see the text). Surfaces, corresponding to $|q(x,y,z)|^2=0.2$, are drawn at $t=141$, $156$, and $177$, the rotation period being $T\approx 48$.

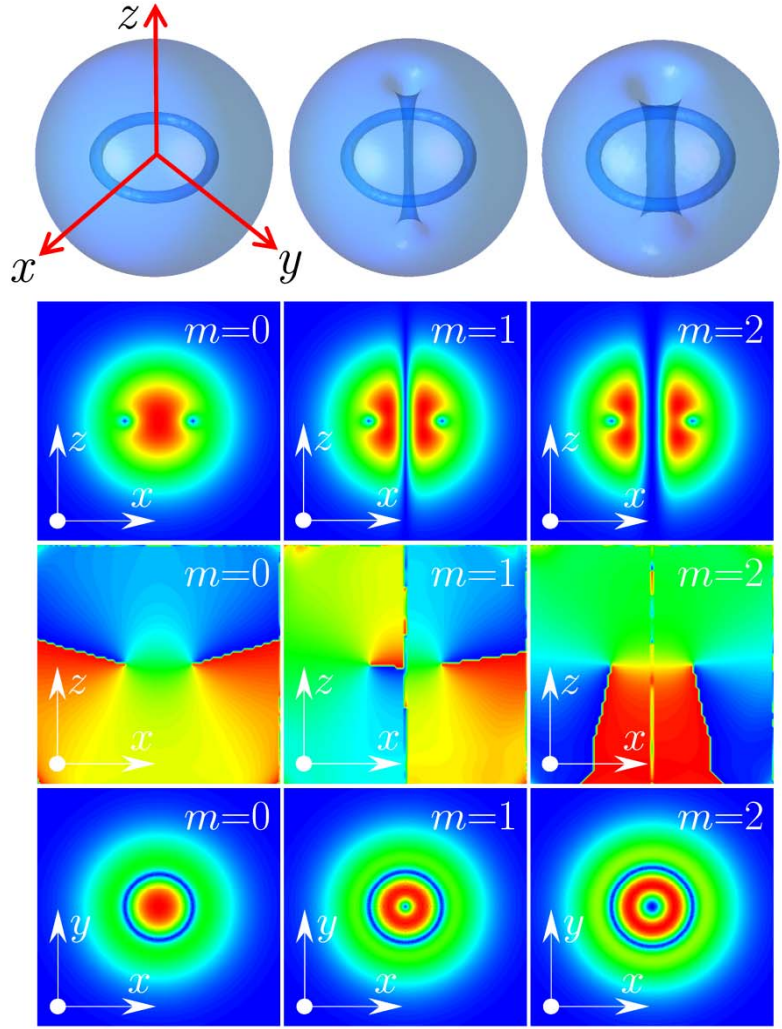

Figure 1. (Color online) Generic examples of "Hopfions", with intrinsic winding number $s=1$ and conventional vorticity $m=0$, $m=1$, or $m=2$, are presented, severally, in the left, central, and right columns, for chemical potential $\mu=15$. The first row shows isosurface density plots at $|q(x,y,z)|^2=1$, the second and third rows display the absolute value and phase of the wave function in cross section $y=0$, and the fourth row shows the distribution of $|q(x,y,z)|$ in the cross-section $z=0$. Here and in other figures, blue and red regions in contour density plots correspond to lower and higher values, respectively.

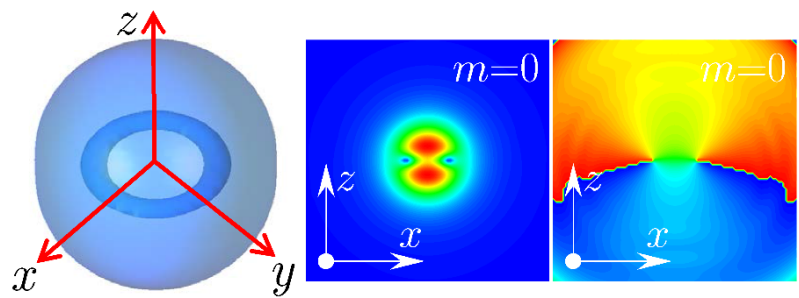

Figure 2. (Color online) Stable "Hopfion" with intrinsic winding number $s=1$ and vorticity $m=0$ at $\mu=10$ supported by algebraic nonlinearity landscape. Isosurface density plot corresponds to $|q(x,y,z)|^2 = 0.4$.

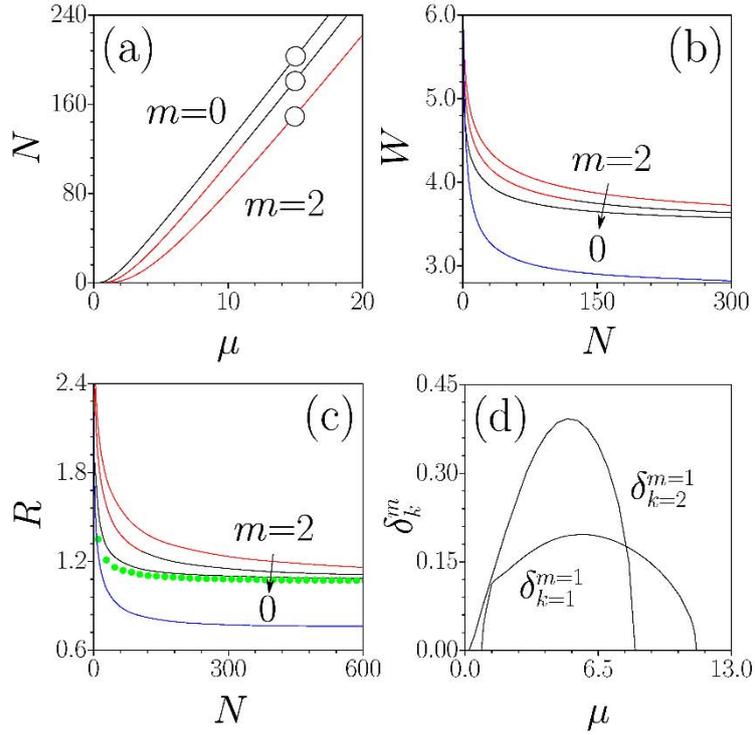

Figure 3. (Color online) Families of twisted toroidal states with intrinsic winding number $s=1$ and different values of the conventional vorticity, $m$: (a) the norm versus the chemical potential; (b) the width versus the norm; (c) the radius of coiled vortex line, $R$, versus the norm. Stable and unstable branches are depicted by black and red segments, respectively. Circles in (a) correspond to the modes displayed in Fig. 1. The chain of green dots in (c) depicts the VA prediction for $R$. Lower blue curves in (b),(c) correspond to $m=0$, $s=1$ solitons in the model with algebraic nonlinearity that are always stable. Due to much more rapid saturation of $R(N)$ dependence for algebraic nonlinearity the norm for blue curve in (c) was multiplied by 3. (d) The real part of the instability growth rate, for different values of azimuthal perturbation index $k$, versus the chemical potential, for $m=1$.

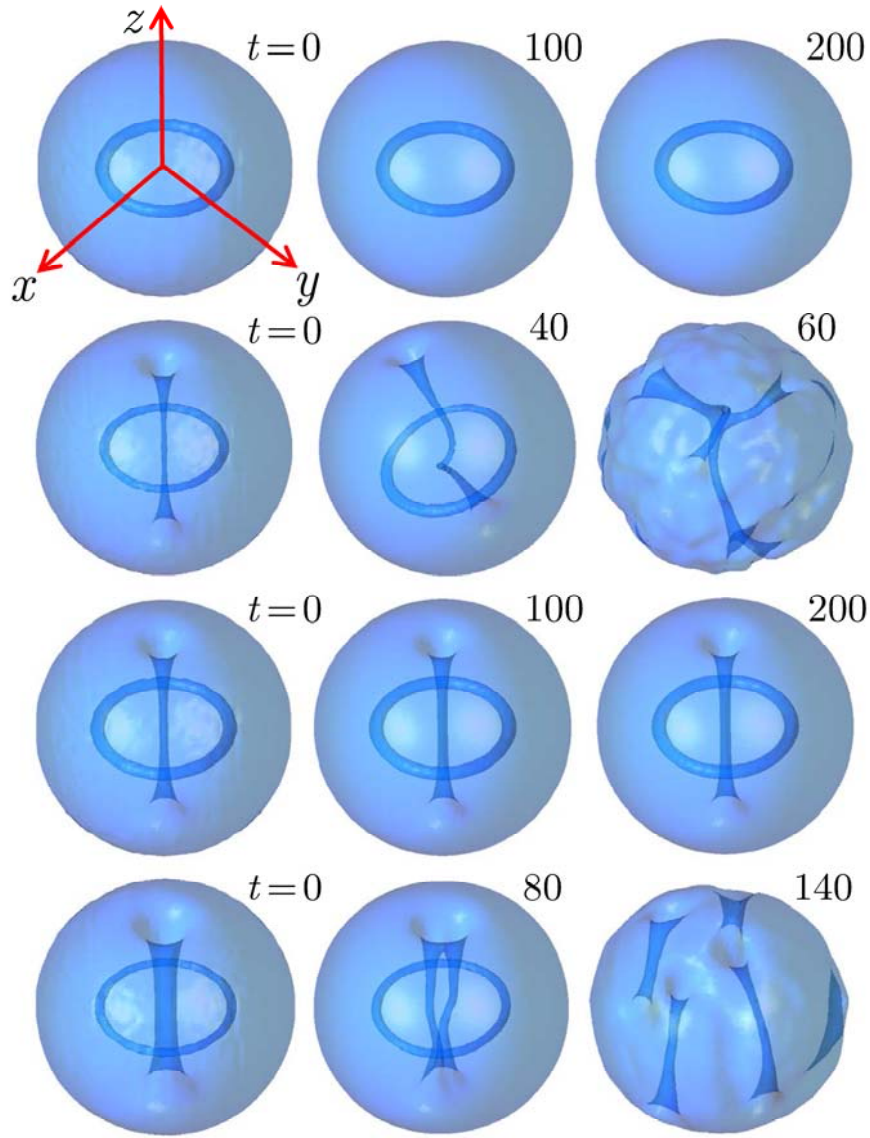

Figure 4. (Color online) Isosurface plots showing the perturbed evolution of the states with $\mu=15$, $m=0$, $s=1$ (the first row), $\mu=9.5$, $m=1$, $s=1$ (the second row), $\mu=16$, $m=1$, $s=1$ (the third row), and $\mu=22$, $m=2$, $s=1$ (the fourth row). Isosurface levels are drawn at $|q(x,y,z)|^2=0.2$ in the second row, and $|q(x,y,z)|^2=1$ in all the others.

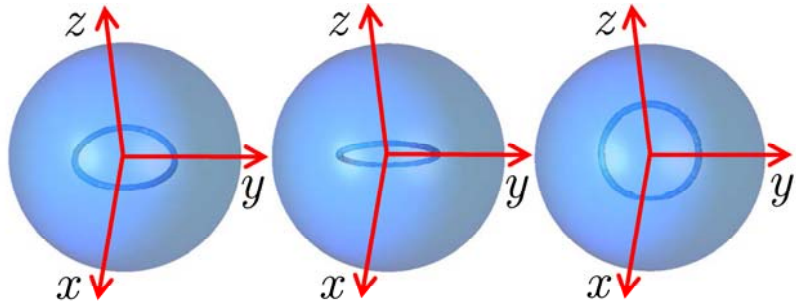

Figure 5. (Color online) Isosurface plots (top) illustrating the solid-ring-like rotation of the simplest twisted toroidal mode, with $m=0$, $s=1$, $\mu=10$, around its diameter (the $y$ axis), initiated by the application of the torque with $\alpha=5$ and $x_0=5$ (see the text). The isosurfaces, corresponding to $|q(x,y,z)|^2=0.2$, are drawn at $t=141$, $156$, and $177$, the rotation period (corresponding to the turn by angle $\pi$) being $T\approx 48$.